\documentclass[journal]{IEEEtran}
\usepackage{cite}

\ifCLASSINFOpdf
  \usepackage[pdftex]{graphicx}
  \graphicspath{{../pdf/}{../jpeg/}}
  \DeclareGraphicsExtensions{.pdf,.jpeg,.png}
\else
  \usepackage[dvips]{graphicx}
  \graphicspath{{../eps/}}
  \DeclareGraphicsExtensions{.eps}
\fi

\usepackage{mhchem}
\usepackage{algorithmic}
\usepackage{array}

\ifCLASSOPTIONcompsoc
  \usepackage[caption=false,font=normalsize,labelfont=sf,textfont=sf]{subfig}
\else
 \usepackage[caption=false,font=footnotesize]{subfig}
\fi

\usepackage{threeparttable}
\usepackage{fixltx2e}
\usepackage{stfloats}
\usepackage{savesym}
\usepackage{amsmath}
\savesymbol{iint}
\usepackage{txfonts}
\restoresymbol{TXF}{iint}

\ifCLASSOPTIONcaptionsoff
  \usepackage[nomarkers]{endfloat}
\let\MYoriglatexcaption\caption
 \renewcommand{\caption}[2][\relax]{\MYoriglatexcaption[#2]{#2}}
\fi

\begin{document}
\title{Critical Current Distributions of Recent Bi-2212 Round Wires}
\author{Shaon Barua,
		Daniel Davis,
		Yavuz Oz,
        Jianyi Jiang,~\IEEEmembership{Senior~Member,~IEEE,}
        Eric~E.~Hellstrom,~\IEEEmembership{Senior~Member,~IEEE,}
        Ulf~P.~Trociewitz,~\IEEEmembership{Senior~Member,~IEEE,}
        and~David~C.~Larbalestier,~\IEEEmembership{Fellow,~IEEE}
\thanks{Manuscript receipt and acceptance dates will be inserted here. The work is supported by the US DOE Office of High Energy Physics under grant number DE-SC0010421 and by the NHMFL, which is supported by NSF under Award Number DMR-1644779, and by the State of Florida, and is amplified by the U.S. Magnet Development Program (MDP). \textit{(Corresponding author: Shaon Barua)}}
\thanks{S. Barua is with the Applied Superconductivity Center, National High Magnetic Field Laboratory, Tallahassee, FL 32310, USA and also with Florida State University (e-mail: sbarua@asc.magnet.fsu.edu).}
\thanks{D. S. Davis, Y. Oz, J. Jiang, E. E. Hellstrom, and U. P. Trociewitz are with National High Magnetic Field Laboratory, Florida State University, Tallahassee, FL 32310, USA.}
\thanks{E. E. Hellstrom, and D. C. Larbalestier are with the National High Magnetic Field Laboratory, Florida State University, Tallahassee, FL 32310, USA and also with the Department of Mechanical Engineering, FAMU-FSU College of Engineering.}
\thanks{Digital Object Identifier 10.1109/TASC.2021.3055479}}

%
%

\markboth{This article has been accepted for publication in a future issue of IEEE Transactions on Applied Superconductivity}%
{Shell \MakeLowercase{\textit{et al.}}: Critical Current Distributions of Recent Bi-2212 Wires}
%



\maketitle

\begin{abstract}
\ce{Bi_{2}Sr_{2}CaCu_{2}O_{8+\text{x}}} (Bi-2212) is the only high-field, high-temperature superconductor (HTS) capable of reaching a critical current density $J_{\text{c}}$(16 T, 4.2 K) of 6500 $\mathrm{A\cdot mm^{-2}}$ in the highly desirable round wire (RW) form. However, state-of-the-art Bi-2212 conductors still have a critical current density ($J_{\text{c}}$) to depairing current density ($J_{\text{d}}$) ratio around 20 to 30 times lower than that of state-of-the-art $\mathrm{Nb-Ti}$ or REBCO. Previously, we have shown that recent improvements in \mbox{Bi-2212} RW $J_{\text{c}}$ are due to improved connectivity associated with optimization of the heat treatment process, and most recently due to a transition to a finer and more uniform powder manufactured by Engi-Mat. One quantitative measure of connectivity may be the critical current ($I_{\text{c}}$) distribution, since the local $I_{\text{c}}$ in a wire can vary along the length due to variable vortex-microstructure interactions and to factors such as filament shape variations, grain-to-grain connectivity variations and blocking secondary phase distributions. Modeling the experimental $V-I$ transition measured on a low resistance shunt as a complex sum of voltage contributions of individual filament and wire sub-sections allows a numerical extraction of the $I_{\text{c}}$ distribution from the $d^{2}V/dI^2$ treatment of the $V-I$ curves. Here we compare $\sim$ 0.1 m length $I_{\text{c}}$ distributions of Bi-2212 RWs with recent state-of-the-art very high-$J_{\text{c}}$ Engi-Mat powder and lower $J_{\text{c}}$ and older Nexans granulate powder.  We do find that the $I_{\text{c}}$ spread for Bi-2212 wires is about twice the relative standard of high-$J_{\text{c}}$ $\mathrm{Nb-Ti}$ well below $H_{\text{irr}}$. We do not yet see any obvious contribution of the Bi-2212 anisotropy to the $I_{\text{c}}$ distribution and are rather encouraged that these Bi-2212 round wires show relative $I_{\text{c}}$ distributions not too far from high-$J_{\text{c}}$ $\mathrm{Nb-Ti}$ wires.
\end{abstract}


%
\IEEEpeerreviewmaketitle

\section{Introduction}

\IEEEPARstart{R}{cent} advances in the critical current density ($J_{\text{c}}$) of powder-in-tube Bi-2212 round wires have generated great interest for their use in high field magnet technology \cite{larbalestier2014isotropic}. An irreversibility field ($H_{\text{irr}}$) of more than 100 T at 4.2~K, along with its macroscopically isotropic, twisted, multi-filamentary architecture make Bi-2212 a promising conductor for applications above 25 T, where the low temperature superconductors (LTS) $\mathrm{Nb-Ti}$ and $\mathrm{Nb_{3}Sn}$ are limited by their much lower irreversibility fields of $\sim$ 11 T and $\sim$ 25 T (at 4.2~K) \cite{miao2012recent}, \cite{parrell2004nb3sn}. Although its $J_{\text{c}}$\mbox{(16 T, ~4.2 K)} = 6500 $\mathrm{A\cdot mm^{-2}}$ far exceeds the Future Circular Collider (FCC) specification of $J_{\text{c}}$(16 T, 4.2 K) = 1500 $\mathrm{A\cdot mm^{-2}}$, $J_{\text{c}}$ for state-of-the-art \mbox{Bi-2212} is still lower than 1\% of the depairing current density \cite{jiang2019high} ($J_{\text{d}} \sim H_{\text{c}}/\lambda \sim 3 \times 10^6$ $\mathrm{A\cdot mm^{-2}}$, where the thermodynamic critical field $H_{\text{c}} \sim 1$ T \cite{kopylov1990role} and the penetration depth ($\lambda$) of \mbox{Bi-2212} $\sim$ 240 nm), as opposed to the 20 – 30\% values achieved with $\mathrm{Nb-Ti}$ or REBCO. Our recent comprehensive survey of many Bi-2212  wires made between 2009 and 2019 has shown that while $J_{\text{c}}$ defined by $I_{\text{c}}/A$, where $A$ is the total Bi-2212 cross-section, varied by almost  a factor of 6 they actually all had almost identical normalized $J_{\text{c}}$($H$), leading us to conclude that very similar vortex pinning properties were shared by all wires and that $|J_{\text{c}}|$ was determined by the effective filament connectivity \cite{brown2019prediction}. 
Filament connectivity is affected by multiple factors on many length scales, including filament cross-section variations before reaction and post-reaction defects such as, cracks, voids, blocking secondary phases, and grain-to-grain connectivity variation. Moreover, cuprate grain boundary $J_{\text{c}}$ drops exponentially with increasing grain-boundary misorientation angle.  Indeed, grain misorientation has historically been the main impediment to the realization of high $J_{\text{c}}$ in cuprate HTSs \cite{larbalestier2014isotropic}. Although the quasi-biaxial texture produced by the RW Bi-2212 heat treatment process is believed to mitigate this high angle grain boundary limitation \cite{kametani2015comparison}, the $c$ - axis rotation of the highly anisotropic Bi-2212 grains results in an inherent $J_{\text{c}}$ variation along the length of each filament. Another commonly observed wire defect is variation of the filament cross section due to the highly non-uniform grain growth characteristics of Bi-2212 from the liquid state. More recently we have seen that the grain texture and filament structure in Bi-2212 RWs are strongly dependent on powder type \cite{jiang2019high}, heat treatment parameters \cite{shen2011heat}, and wire diameter \cite{jiang2016effects}.

The $I_{\text{c}}$ distribution is a potential quantitative measure of this complex connectivity variation. We are motivated by the possibility that this distribution can identify avenues for improving $I_{\text{c}}$ and perhaps for wire production quality control. Recently, a transition to finer powder made by Engi-Mat has more than doubled $J_{\text{c}}$ in state-of-the-art Bi-2212 wires \cite{jiang2019high} compared to earlier wires made with Nexans granulate powder. To extract the $I_{\text{c}}$ distribution of our wires, we employed the method previously employed to characterize the $I_{\text{c}}$ distribution from $V-I$ characteristics of  $\mathrm{Nb-Ti}$ and $\mathrm{Nb_{3}Sn}$ wires using a normal shunt to make $d^{2}V/dI^2$ analysis possible \cite{BAIXERAS19671541, warnes1986critical, warnes1988model, mueller2007critical}. We should note that the shunt resistance of the normal metal matrices of Cu in $\mathrm{Nb-Ti}$ and Ag in Bi-2212 are rather similar, both having residual resistivity ratio (RRR) $>$ 100 \cite{li2015rrr, bonura2018very}. 
\begin{table*}[t]
  \centering
  \caption{Sample Specifications}
    \begin{tabular}{lllllllll}\hline\hline
    Sample & Material & \multicolumn{1}{l}{Powder} & Sheath & Manufacturer & \multicolumn{1}{l}{Diameter} & \multicolumn{1}{l}{No. of} & \multicolumn{1}{l}{Filling } & \multicolumn{1}{l}{$J_{\text{c}}$(5 T, 4.2 K)} \\
          &       &       & Material &       & \multicolumn{1}{l}{[mm]} & \multicolumn{1}{l}{Filaments} &   {Factor}    & \multicolumn{1}{l}{[$\mathrm{A\cdot mm^{-2}}$]} \\\hline
    pmm180410 & Bi-2212 & \multicolumn{1}{l}{Engi-Mat} & Ag--0.2 wt.\% Mg & B-OST & 1.0     & 85$\times$18    & 0.201 & 7115 \\
    pmm100913 & Bi-2212 & \multicolumn{1}{l}{Nexans} & Ag--0.2 wt.\% Mg  & B-OST & 0.8   & 37$\times$18    & 0.221 & 4085 \\
    Nb – Ti & Nb--47 wt.\% Ti &       & Cu    & Supercon Inc. & 0.6   & 54    & 0.436 & 3280 \\\hline\hline
    \end{tabular} 
  \label{tab:addlabel}%
\end{table*}%

\section{Experimental}
Two Bi-2212 RWs with the same nominal powder composition of Bi$_{2.17}$Sr$_{1.94}$Ca$_{0.90}$Cu$_{1.98}$O$_{\text{x}}$ but very different powders and time of fabrication were selected. The pmm100913 and pmm180410 wires were fabricated in 2010 and 2018 by Bruker-Oxford Superconducting Technology (B-OST) using Nexans granulate (lot 77) and Engi-Mat (LXB 116) powder.  Both powders were state-of-the-art at the time. Double restack multifilamentary wires were made by the Powder-In-Tube (PIT) process using powder-filled pure Ag tubes and an outer sheath of Ag--0.2 wt.\% Mg alloy. The heat treatment of 1.5~m spiral wires was done at 50 bar over-pressure with an oxygen partial pressure of 1 bar \cite{jiang2019high}.  Wires were then soldered onto a brass ITER-like barrel\cite{bruzzone1996bench} and the $V-I$ characteristics of multiple 10 cm sections of each sample were measured using standard four-probe transport measurement methods at 4.2 K in magnetic fields up to 15 T. The peak applied currents were well above the standard 1 $\muup\mathrm{V\cdot cm^{-1}}$ criterion limit far into the flux flow regime, so as to obtain the full $I_{\text{c}}$ distribution. A standard Nb--47 wt.\% Ti wire of high $J_{\text{c}}$ manufactured by Supercon Inc. was also measured. Detailed specifications of the samples are listed in Table I.

\subsection{Measurements of the Critical Current Distribution }
Traditional four-probe transport measurements reveal only the lower end of the $I_{\text{c}}$ distribution. In a well-optimized material like $\mathrm{Nb-Ti}$ where current is believed to flow uniformly, magnetization and transport measurements typically agree well when corrections for electric field ($E$) differences are made (and filament sausaging is avoided). Bi-2212 is more complex: the current path is percolative and uncertain with a fractional occupancy of as little as 1\% \cite{brown2019prediction}. Moreover, most transport measurements typically characterize only $\sim$ 1 cm of conductor due to use of straight wires in narrow-bore magnets. Because of the greater stability of Bi-2212 compared to $\mathrm{Nb-Ti}$ and $\mathrm{Nb_{3}Sn}$ it appears that a relatively high criterion of 1 $\muup\mathrm{V\cdot cm^{-1}}$ reliably predicts the quench performance of even potted magnets \cite{shen2019stable}, making it interesting to explore the $V-I$ transition at both lower and higher $E$ values. 

An early method that was successfully used to characterize the $I_{\text{c}}$ distribution of $\mathrm{Nb-Ti}$ and $\mathrm{Nb_{3}Sn}$ wires used a substantial normal metal shunt to avoid conductor burn out \cite{warnes1986critical, plummer1987dependence}. The method models the $I_{\text{c}}$ distribution as a series array of many short longitudinal sections of varying $I_{\text{c}}$ \cite{mueller2007critical, kimmich1999investigation}. As the applied current exceeds the local critical current ($i_\text{c}$) of a section, the section transitions from a pinned to a flux flow state. Since the flux flow resistivity is two to three orders of magnitude higher than the silver matrix and the high conductance barrel\cite{warnes1986critical}, almost all the excess current is carried by the normal matrix and shunt without excessive heating and depinning of vortices in the superconductor. As originally formulated by Baixeras and Fournet\cite{BAIXERAS19671541}, the voltage across a sub-element with critical current $i_{\text{c}}$ as a function of applied current $I$ is given by a series distribution of variable $i_{\text{c}}$ elements:

\begin{equation}
V(I,i_\text{c}) = R(I-i_\text{c})
\end{equation}

The normal method of degradation of $I_{\text{c}}$ in $\mathrm{Nb-Ti}$ is by onset of filament sausaging which fits the series $i_{\text{c}}$ array model well but almost all other practical superconductors have more complex series-parallel current paths.  Plummer and Evetts\cite{plummer1987dependence}  extended this analysis to filamentary $\mathrm{Nb_{3}Sn}$ conductors, where the normal bronze matrix can have higher ohmic dissipation than  flux flow in the $\mathrm{Nb_{3}Sn}$ filaments (bronze Cu is always poisoned by the residual Sn content, typically 0.15 – 1 at.\% meaning that its RRR is $<$ 5), using a dynamic flux flow relation to determine the $\mathrm{Nb_{3}Sn}$ distribution, arriving at the same equation as Baixeras and Fournet. In their model, the $I_{\text{c}}$ distribution was assumed to be Gaussian due to many variations of grain size, local Sn content and filament uniformity, none of which were well controlled or easy to measure. In this more general case of a wire with many filaments, varying vortex pinning and varying active cross-section, the distribution of $i_{\text{c}}$ values is both parallel and series and complex. Denoting the desired distribution $\phi(i_{\text{c}})$  as the probability distribution density of $i_{\text{c}}$ in the wire suggests use of a Gaussian function since many independent factors are locally determining the local $i_{\text{c}}$ values. Taking $R_{\text{eff}}$ as the effective resistance of the normal currents in the stabilizer and shunt, integrating over all possible $i_{\text{c}}$ values gives the following expression for the total voltage measured across the wire at a given applied current $I$:
\begin{equation}
V(I)=R_\text{eff}\int_{0}^{I}(I-i_\text{c})\phi(i_\text{c})di_\text{c}
\end{equation}
\begin{equation}
\phi(I)=\frac{1}{\sqrt{2\pi\sigma^2}}\exp\frac{-(I-\mu)^2}{2\sigma^2}
\end{equation}
\begin{equation}
\frac{d^{2}V(I)}{dI^2}=R_\text{eff}\phi(I)= \frac{R_\text{eff}}{\sqrt{2\pi\sigma^2}}\exp\frac{-(I-\mu)^2}{2\sigma^2}
\end{equation} 

Differentiating $V(I)$ twice with respect to $I$ yields $R_{\text{eff}}\cdot \phi(I)$, which allows extraction of the $I_{\text{c}}$ distribution and $R_{\text{eff}}$. To reduce noise, a Savitzky-Golay filter was used to smooth the data before calculating the second derivative\cite{warnes1986critical, savitzky1964smoothing} by regression fitting a second order polynomial to a typical comb size ($m$) of 7 – 15 $V-I$ points, the analytical derivative being determined from the coefficient at the center of a group of $2m + 1$ points. The 7 – 15 point comb size was kept low to avoid smoothing out subtle features of the distribution curve shape.  Since the factors that affect local critical current are thought to be randomly distributed and uncorrelated, the distribution of $i_{\text{c}}$ plausibly converges to a Gaussian as predicted by the central limit theorem\cite{mueller2007critical, plummer1987dependence, edelman1993resistive, hampshire1987detailed} enabling extraction of distribution parameters, such as the mean $I_{\text{c}}$ ($\mu$ in (4)) and the standard deviation ($\sigma$).

\subsection{The Fraction of Superconductor in Flux Flow}
$f_{\text{D}}(I)$ represents the fraction of superconductor wire in the flux flow state at a given current and this was calculated from $d^{2}V/dI^2$ at $I_{\text{c}}$ determined by both 0.1 $\muup\mathrm{V\cdot cm^{-1}}$ and 1~$\muup\mathrm{V\cdot cm^{-1}}$ criteria according to:
\begin{equation}
f_\text{D}(I)=\int_{0}^{I}\phi(i_\text{c})di_\text{c}
\end{equation}

\begin{figure}
\centering
	\includegraphics[width=\linewidth]{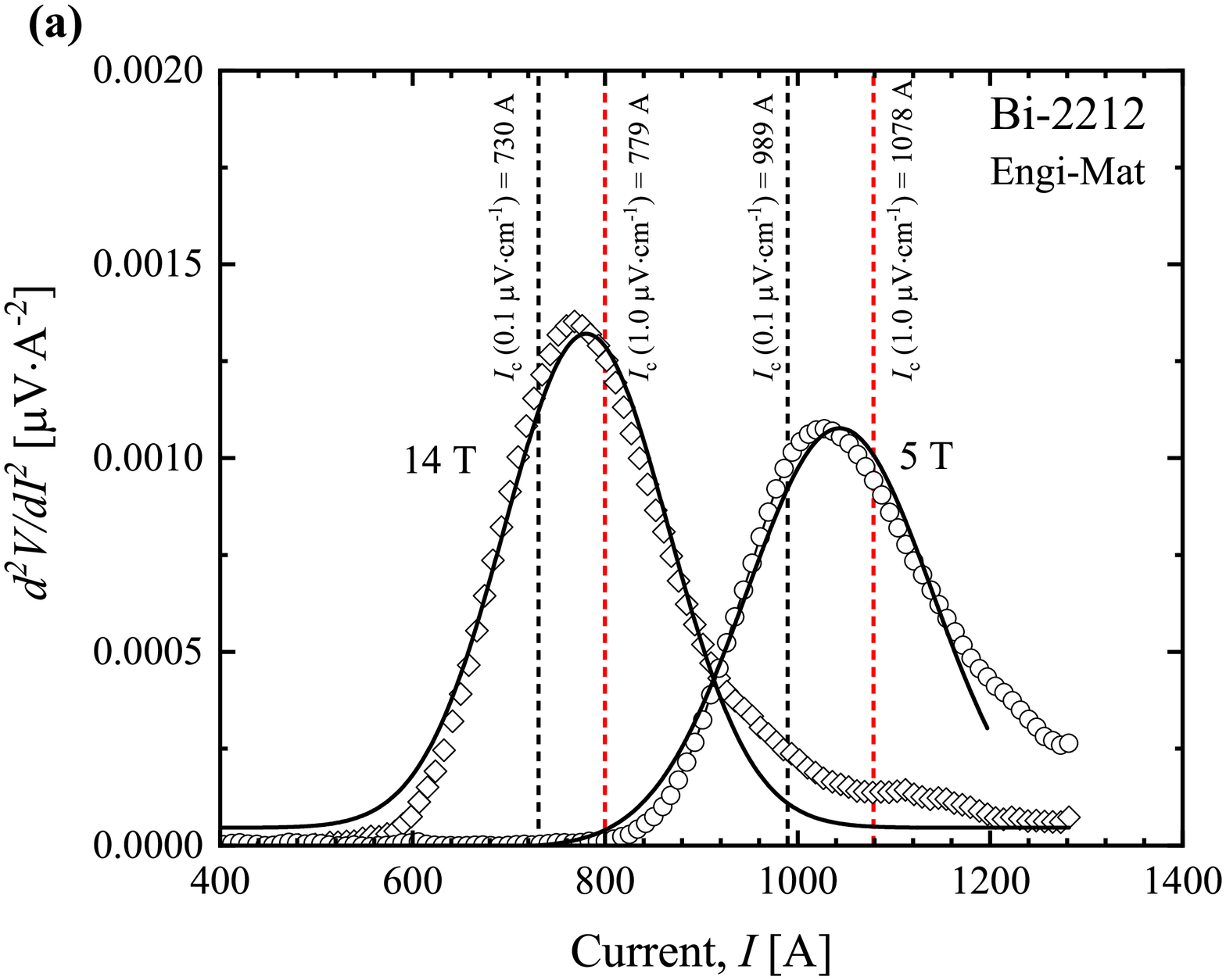}
	\includegraphics[width=\linewidth]{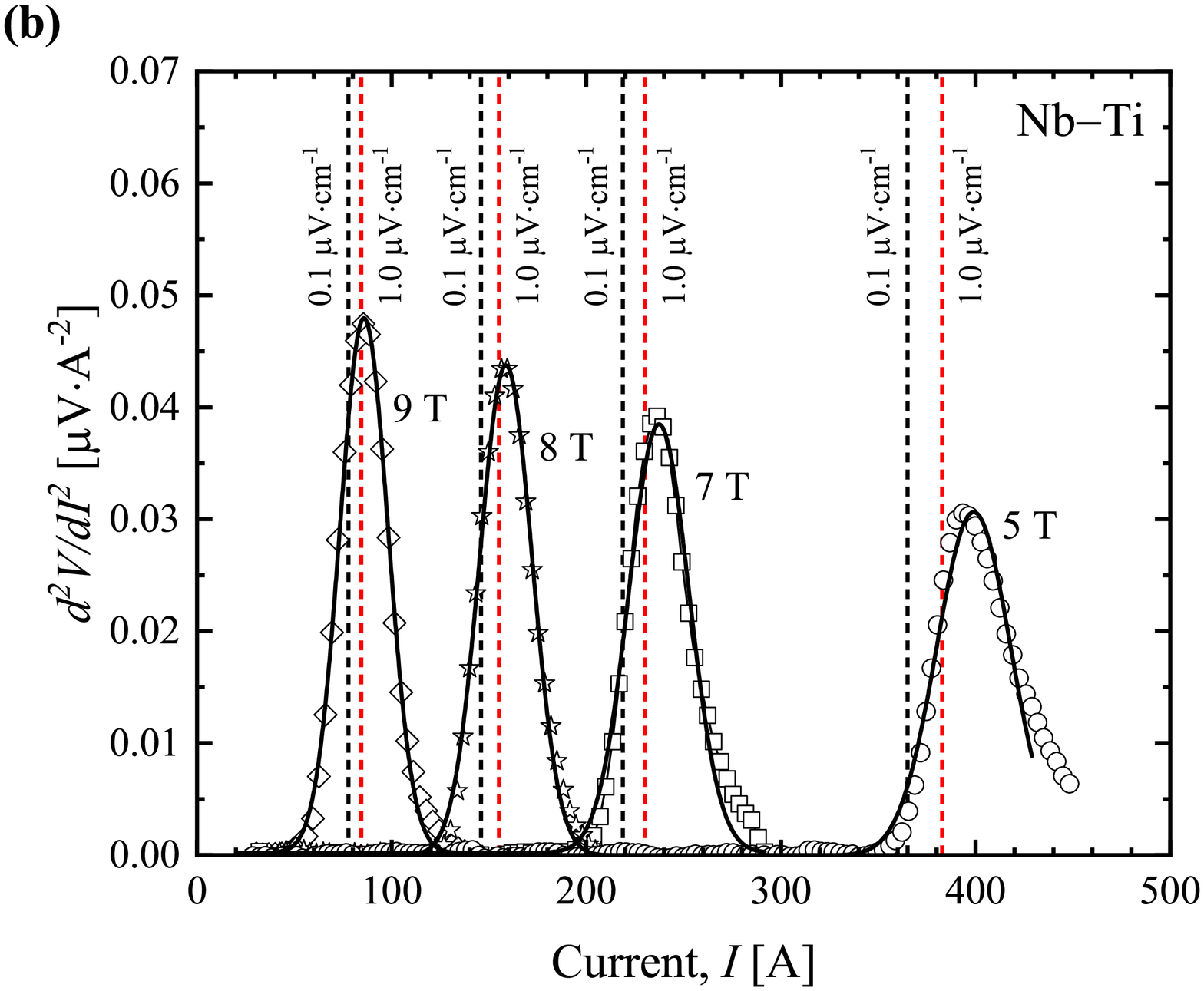}
	\caption{Critical current distribution of (a) pmm180410 (Engi-Mat) wire at \mbox{5 T} (circle) and 14 T (diamond) at 4.2 K, (b) $\mathrm{Nb-Ti}$ wire at 5 T (circle), 7 T (square), 8 T (star), and 9 T (diamond) at 4.2 K. Two separate lines are drawn in the distribution plot for each field to delineate the short sample $I_{\text{c}}$ based on 0.1 $\muup\mathrm{V\cdot cm^{-1}}$ (black dotted line) and 1 $\muup\mathrm{V\cdot cm^{-1}}$ (red dotted line) criterion. Solid black lines represents the Gaussian distribution.}
\label{Fig.}
\end{figure}
\subsection{Critical Current Measurements of Short (4.5 cm) Samples}
To compare regular measurements on short samples to those determined on the shunted barrel samples, 10 cm long wires of Bi-2212 were heat treated along with each spiral sample. Transport critical currents $I_{\text{c}}$ were measured using the four-probe method at 4.2 K in perpendicular magnetic fields up to 15 T using both 0.1 $\muup\mathrm{V\cdot cm^{-1}}$ and 1~$\muup\mathrm{V\cdot cm^{-1}}$ criteria.  The resistive transition index ($n$ value) was calculated by fitting the $V-I$ curve from 0.6 – 20 $\muup\mathrm{V\cdot cm^{-1}}$.

\begin{figure}[t]
\centering
	\includegraphics[width=\linewidth]{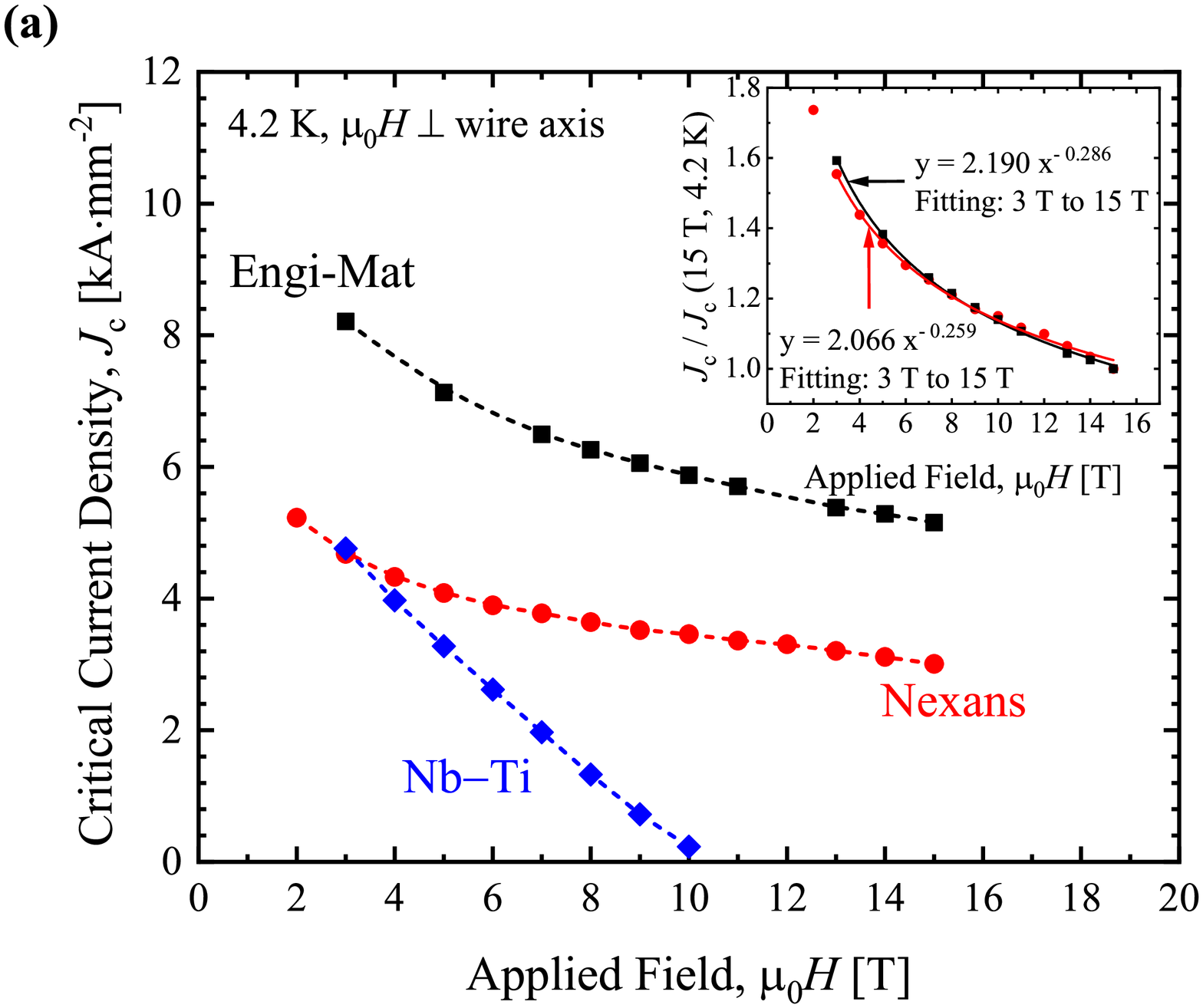}

	\includegraphics[width=\linewidth]{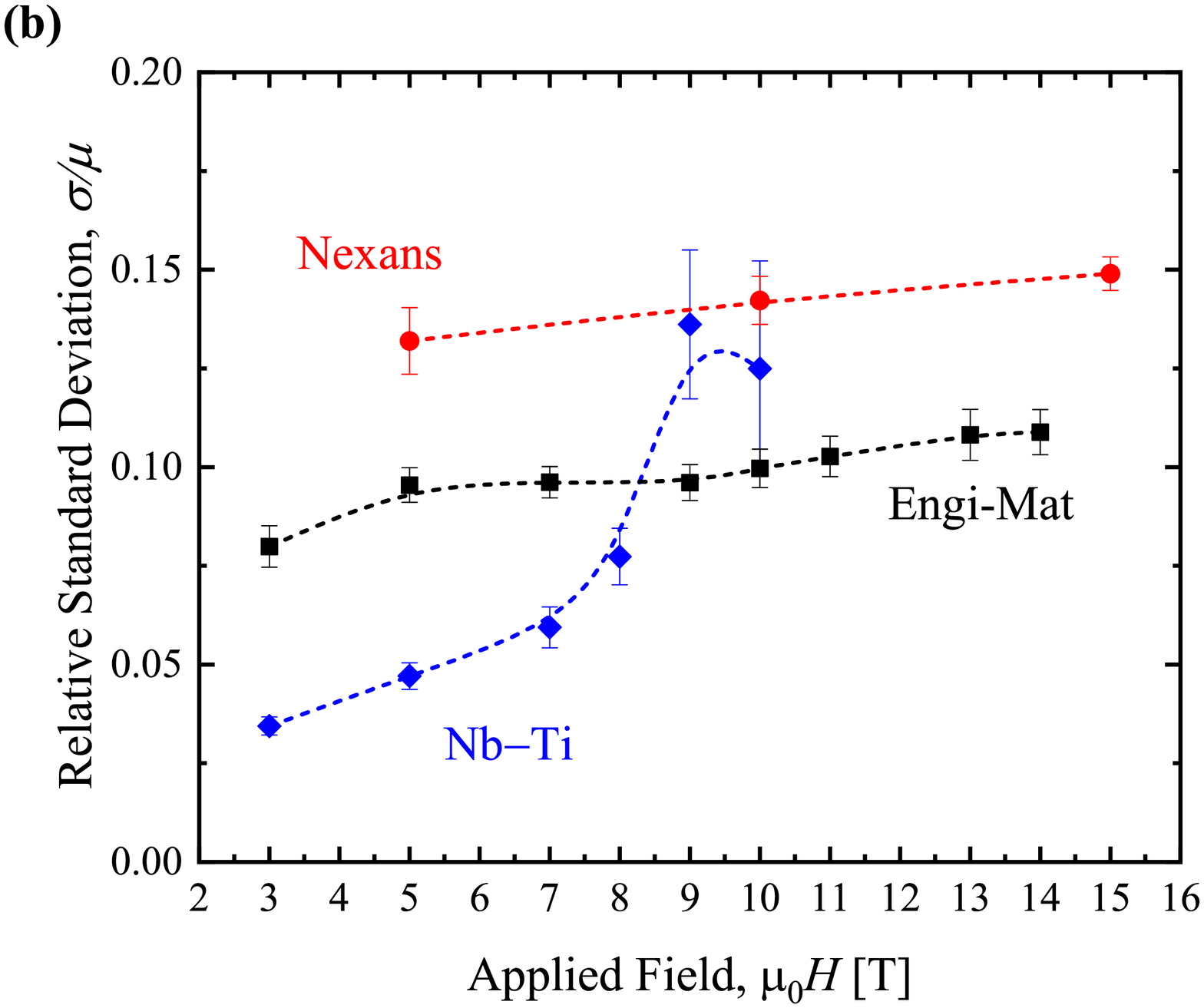}
	\caption{(a) Critical current density evaluated at 1 $\muup\mathrm{V\cdot cm^{-1}}$, and (b) relative standard deviation of $I_{\text{c}}$ distribution as a function of applied field for pmm180410 (square), pmm100913 (circle), and $\mathrm{Nb-Ti}$ (diamond) wires at 4.2 K. The power law fit from 3 T to 15 T for both Bi-2212 wire is shown in the inset. Dashed lines are guides for the eye.}

\label{Fig.}
\end{figure}

\section{Results}
Fig. 1(a) shows the $I_{\text{c}}$ distribution of a 10 cm section of the higher-$J_{\text{c}}$ Engi-Mat wire at 5 T and 14 T. The basic features are similar in all measured samples. Both 5 T and 14 T curves are slightly asymmetric due to an extended high current tail but it is also evident that the 14 T $I_{\text{c}}$ distribution is sharper than the 5 T distribution. Two separate lines mark the short sample $I_{\text{c}}$ position based on 0.1 $\muup\mathrm{V\cdot cm^{-1}}$ and 1 $\muup\mathrm{V\cdot cm^{-1}}$ criteria. The $I_{\text{c}}$ of the higher-$J_{\text{c}}$ Engi-Mat short sample at 1~$\muup\mathrm{V\cdot cm^{-1}}$ is 1078 A at 5 T and 782 A at 14 T compared to the distribution analysis mean $I_{\text{c}}$ $\sim$ 1039.8 $\pm$ 16.9 A, $\sigma \sim$ 99.3 $\pm$ 4.9 A, and $\sigma/\mu$ $\sim$ 0.095 $\pm$ 0.0044 at 5 T. By contrast, the older and lower-$J_{\text{c}}$ Nexans wire shows mean $I_{\text{c}}$ $\sim$ 448.2 $\pm$ 5.1 A, $\sigma \sim$ 59.1 $\pm$ 3.4 A, and  $\sigma/\mu$ $\sim$ 0.132 $\pm$ 0.0084 at 5 T. For comparison to a more classical wire Fig. 1(b) shows the $I_{\text{c}}$ distribution of the Supercon $\mathrm{Nb-Ti}$ wire at various fields. At 5 T, it has $I_{\text{c}}$ $\sim$ 394.9 $\pm$ 3.6 A, $\sigma \sim$ 18.6 $\pm$ 1.5 A, and $\sigma/\mu$ $\sim$ 0.047 $\pm$ 0.0034. The $\mathrm{Nb-Ti}$ wire distribution width is significantly lower than both Bi-2212 samples at fields well below $H_{\text{irr}}$ but does cross the Engi-Mat Bi-2212 wire close to $H_{\text{irr}}$. Further details of the distribution parameters are listed in Table II. 

Fig. 2(a) shows $J_{\text{c}}(H)$ for Engi-Mat, Nexans and $\mathrm{Nb-Ti}$ wires. The power law fit of $J_{\text{c}}(H)$ from 3 T to 15 T for both Bi-2212 wires is shown in the inset of Fig. 2(a).

Fig. 2(b) shows $\sigma/\mu$ of the $I_{\text{c}}$ distribution for Engi-Mat, Nexans, and $\mathrm{Nb-Ti}$ wires with error bars of one standard deviation calculated from multiple sections of the spiral on each barrel. For $\mathrm{Nb-Ti}$, $\sigma/\mu$ increases monotonically with field from 0.034 $\pm$ 0.0023 at 3 T to 0.125 $\pm$ 0.0272 at 10~T, but it varies only slowly with increasing magnetic field in \mbox{Bi-2212} wires, most likely because, unlike for $\mathrm{Nb-Ti}$, $H/H_{\text{irr}}$ is always significantly lower than unity. At 10 T, Nexans and Engi-Mat wires show $\sigma/\mu$ $\sim$ 0.14, and 0.10, respectively, but is then 0.13 for $\mathrm{Nb-Ti}$. The fractional energy dissipation ($f_{\text{D}}$) and the ratio of short sample critical current to mean critical current ($I_{\text{c}}/\mu$) are listed in Table II. The $f_{\text{D}}$($I_{\text{c}}$) of the Engi-Mat wire, where $I_{\text{c}}$ using the 1 $\muup\mathrm{V\cdot cm^{-1}}$ electric field criterion on the 4.5 cm long sample shows that 60 – 65\% of the superconductor has transitioned from the flux pinning to the flux flow state. In contrast, the more conservative 0.1~$\muup\mathrm{V\cdot cm^{-1}}$ criterion corresponds to 22 – 27\% in flux flow. In the lower-$J_{\text{c}}$ Nexans wire $f_{\text{D}}$ ranges from 43 – 50\% at $I_{\text{c}}$ (1~$\muup\mathrm{V\cdot cm^{-1}}$), while the conservative ($I_{\text{c}}$ 0.1 $\muup\mathrm{V\cdot cm^{-1}}$) criterion yields $\sim$ 15 – 19\%. For the $\mathrm{Nb-Ti}$ wire, $f_{\text{D}}$ increased from 14\% to 69\% with an increasing magnetic field.  The $I_{\text{c}}/\mu$ ratio of Engi-Mat, Nexans, and $\mathrm{Nb-Ti}$ wires are $\sim$ 1.04, $\sim$ 0.98 – 1.0, and 0.94 – 1.07, respectively.

\begin{table*}[h]
  \centering
  \caption{Parameters Derived From The Critical Current Distribution}
  \begin{threeparttable}
    \begin{tabular}{lllllllllll} \hline\hline
    \multicolumn{1}{l}{Sample\tnote{\textdagger}} & \multicolumn{1}{l}{Magnetic} & \multicolumn{1}{l}{Mean, $\mu$} & \multicolumn{1}{l}{Standard} & \multicolumn{1}{l}{Relative Std.} &  $R_\text{eff}\tnote{+}$     & \multicolumn{1}{l}{Critical} & \multicolumn{1}{l}{$n$ value} & \multicolumn{1}{l}{$f_{D}(I_\text{c})$\tnote{a}} & \multicolumn{1}{l}{$f_\text{D}(I_\text{c})$\tnote{b}} & $ \frac{I_\text{c}\tnote{++}} {\mu}$ \\
          & \multicolumn{1}{l}{Field} &       & \multicolumn{1}{l}{Deviation, $\sigma$ } & \multicolumn{1}{l}{Deviation, $\sigma/\mu$} &       & \multicolumn{1}{l}{Current, $I_\text{c}$\tnote{++}} &       &       &       &  \\
          &       &       &       &       &       &       &       &       &       &  \\
          &       &       &       &       &       &       &       &       &       &  \\
          & \multicolumn{1}{l}{[T]} & \multicolumn{1}{l}{[A]} & \multicolumn{1}{l}{[A]} &       & \multicolumn{1}{l}{$\mathrm{[\muup\Omega\cdot cm^{-1}]}$} & \multicolumn{1}{l}{[A]} &       & \multicolumn{1}{l}{[\%]} & \multicolumn{1}{l}{[\%]} &  \\\hline
              
    \multicolumn{1}{l}{pmm180410} & 3     & 1210.6 & 96.6  & 0.080 & 0.023 & 1242.5 & 26.6  & 22.8  & 61.9  & 1.03 \\(Engi-Mat) & 5     & 1039.8 & 99.3  & 0.095 & 0.028 & 1078.8 & 26.7  & 30.5  & 64.2  & 1.04 \\
          & 7     & 947.6 & 91.2  & 0.096 & 0.027 & 982.6 & 25.6  & 28.7  & 64.1  & 1.04 \\
          & 9     & 881.3 & 84.4  & 0.096 & 0.027 & 916.5 & 26.2  & 29.6  & 63.7  & 1.04 \\
          & 10    & 859.6 & 85.7  & 0.100 & 0.028 & 888.9 & 25.7  & 29.5  & 63.1  & 1.03 \\
          & 11    & 834.0 & 85.7  & 0.103 & 0.028 & 863.3 & 25.2  & 29.1  & 62.6  & 1.04 \\
          & 13    & 793.8 & 85.9  & 0.108 & 0.029 & 814.1 & 25.0    & 24.9  & 60.9  & 1.03 \\
          & 14    & 774.2 & 84.3  & 0.109 & 0.029 & 799.7 & 22.2  & 30.6  & 62.1  & 1.03 \\
          &       &       &       &       &       &       &       &       &       &  \\
    \multicolumn{1}{l}{pmm100913} & 5     & 448.2 & 59.1  & 0.132 & 0.030  & 445.2 & 17.4  & 16.5  & 49.6  & 0.99 \\
          (Nexans)& 10    & 376.6 & 53.5  & 0.142 & 0.032 & 377.2 & 18.8  & 18.6  & 47.8  & 1.00 \\
          & 15    & 336.2 & 50.1  & 0.149 & 0.032 & 328.0 & 15.2  & 13.3  & 43.8  & 0.98 \\
          &       &       &       &       &       &       &       &       &       &  \\
    \multicolumn{1}{l}{Nb – Ti} & 3     & 589.7 & 20.3  & 0.034 & 0.047 & 556.1 & 50.2  & 1.7   & 14.3  & 0.94 \\
          & 5     & 394.9 & 18.6  & 0.047 & 0.048 & 382.8 & 62.6  & 8.1   & 30.5  & 0.97 \\
          & 7     & 233.0 & 13.8  & 0.059 & 0.048 & 230.1 & 58.6  & 18.6  & 43.9  & 0.99 \\
          & 8     & 154.9 & 12.0  & 0.077 & 0.048 & 155.2 & 55.2  & 23.6  & 53.5  & 1.00 \\
          & 9     & 82.1  & 11.2  & 0.136 & 0.048 & 84.4  & 44.5  & 36.1  & 59.9  & 1.03 \\
          & 10    & 25.1  & 3.1   & 0.125 & 0.049 & 26.9  & 15.8  & 21.3  & 68.5  & 1.07 \\ \hline\hline
    \end{tabular}
\begin{tablenotes}
    \item[\textdagger]Distribution is measured for multiple sections of the spiral sample at each field. Here average results of the fitting parameters are listed. 
    \item[+]Each voltage tap in pmm180410, pmm100913, and $\mathrm{Nb-Ti}$ samples is 10 cm, 20 cm, and 30 cm apart, respectively. 
    \item[++]Critical current is calculated from a 4.5 cm short sample based on 1 $\muup\mathrm{V\cdot cm^{-1}}$ criterion at 4.2 K.  
    \item[a]$f_{\text{D}}(I_{\text{c}})$ is calculated based on short sample $I_{\text{c}}$ at 0.1 $\muup\mathrm{V\cdot cm^{-1}}$ criterion.
    \item[b]$f_{\text{D}}(I_{\text{c}})$ is calculated based on short sample $I_{\text{c}}$ at 1 $\muup\mathrm{V\cdot cm^{-1}}$ criterion.
\end{tablenotes}
    \end{threeparttable}
  \label{tab:addlabel}%
\end{table*}%

 \section{Discussion}
In this work, we have compared the $I_{\text{c}}$ distribution of two \mbox{Bi-2212} RWs made by B-OST prepared with newer and better and older and worse quality powders from Engi-Mat and Nexans. We were motivated to better understand the conclusion of Brown \textit{et al.} \cite{brown2019prediction} that the magnitude of $J_{\text{c}}$ in Bi-2212 wires was almost completely determined by the effective connectivity of the current path which we interpreted to be very small since the ratio $J_{\text{c}}/J_{\text{d}}$ is of order 1\% or less. Even if the vortex pinning in Bi-2212 is weak (it is also not yet clarified and may just be due to cation defect fluctuations within the Bi-2212 structure), such a low ratio of $J_{\text{c}}$ to $J_{\text{d}}$ implies that the long range current path may occupy a small fraction of the total cross-section. In the absence of detailed measurements of $J_{\text{c}}$ at the filament level (we have done this earlier for sections of Bi-2223 filament and found values of $J_{\text{c}}$ several times the average\cite{cai1998current}), we started this work to see if the distribution of $J_{\text{c}}$ values extracted from $d^{2}V/dI^2$ measurements would offer a better understanding of the differences between representative wires made with the former champion Nexans and the more recent champion Engi-Mat powders with about a factor of 2 difference in the filament $J_{\text{c}}$ defined by the measurement of $I_{\text{c}}$ divided by the fully densified Bi-2212 cross-section.

In analyzing the data we take the common view \cite{warnes1988model, edelman1993resistive, hampshire1987detailed, jones1967non} that the local $I_{\text{c}}$ along the wire is controlled by multiple independent factors that justify a Gaussian approximation consistent with the central limit theorem \cite{plummer1987dependence}. Gaussian distributions have earlier been measured in $\mathrm{Nb_{3}Sn}$ \cite{mueller2007critical, plummer1987dependence, kimmich1999investigation} with $\sigma/\mu$ values of 0.15 at 15 T, about 0.6 $H_{\text{irr}}$. Here we show that a high-$J_{\text{c}}$ $\mathrm{Nb-Ti}$ has a significantly smaller $\sigma/\mu$ of $\sim$ 0.05 at 5 T ($\sim$ 0.5 $H_{\text{irr}}$), although it does rise significantly above 0.1 at 8 T close to $H_{\text{irr}}$. Both $\mathrm{Nb-Ti}$ and $\mathrm{Nb_{3}Sn}$ are isotropic with respect both to vortex pinning and to $H_{\text{c2}}$ \cite{suenaga1991irreversibility, brandt1995flux}. Given the meandering $c$ - axis along the filament, we expect that some grains will have very favorable orientations for $H_{\text{c2}}$ and some quite unfavorable. Not yet being clear what the strong pinning centers are in Bi-2212, we cannot yet \textit{a priori} say whether the high $H_{\text{c2}}$ $ab$ - plane orientation has stronger pinning than the $c$ - axis orientation but we might expect that the $d^{2}V/dI^2$ would be broadened on the high $I_{\text{c}}$ side by this anisotropy.  

Indeed, we found that our two Bi-2212 wires showed Gaussian behavior over about three quarters of their $I_{\text{c}}$ distribution but also possessed a notable non-Gaussian extended tail beyond about 150\% of the mean $I_{\text{c}}$. Such a tail was also seen in studies of Bi-2223 tapes \cite{hornung2010current}. Since both Bi-2223 and Bi-2212 are strongly anisotropic cuprates and both have percolative current paths, broader transitions and some non-Gaussian behavior is not unexpected, particularly at higher $I_{\text{c}}$ values that may correspond to grains with higher $H_{\text{c2}}$ values. Qualitatively then the non-Gaussian high-side tail has a plausible physical basis, even if we are very far from being able to access the whole range of $i_{\text{c}}$ of grains with strongly varying orientations.

It is worth pointing out that the Bi-2212 distributions yield mean critical current values $\mu$ within a few percent of the measured short ($\sim$ 5 cm) sample $I_{\text{c}}$(1 $\muup\mathrm{V\cdot cm^{-1}}$) results, with a dissipation fraction $f_{\text{D}}$ rather independent of field, whereas for $\mathrm{Nb-Ti}$ $f_{\text{D}}$ increases from 15\% to 68\% between 3 T and 10~T with $\mu$ moving from above to below $I_{\text{c}}$.  This corresponds to a decrease in dissipation from $\sim$ 200 $\mathrm{mW\cdot cm^{-3}}$ at 3 T to only 10 $\mathrm{mW\cdot cm^{-3}}$ at 10 T at a 1 $\muup\mathrm{V\cdot cm^{-1}}$ criterion, in the context of the low enthalpy margin of $\mathrm{Nb-Ti}$, on the order of 1 – 10 $\mathrm{mJ\cdot cm^{-3}}$. The higher dissipation occurring at lower fields causes LTS conductors to quench rapidly when even a fraction of the distribution is normal, as seen by the higher $n$ values of their transitions. By contrast, for Bi-2212, the dissipation changes are much less, $\sim$ 160 – 100 $\mathrm{mW\cdot cm^{-3}}$ from 3 – 14 T for the Engi-Mat wire, and the enthalpy margin is two orders of magnitude larger, leading to more uniform $f_{\text{D}}$ and $n$ values. At higher fields the margin decays only gradually for Bi-2212, which still has over 10 times more margin at 30 T than $\mathrm{Nb-Ti}$ has at 5 T.

The relative standard deviation ($\sigma/\mu$) is a significant parameter which considers both the breadth and the mean of the $I_{\text{c}}$ distribution. Previous investigations of the $I_{\text{c}}$ distribution of technological conductors yielded $\sigma/\mu$ $\sim$ 0.02 – 0.50 for $\mathrm{Nb_{3}Sn}$, $\mathrm{Nb-Ti}$, and Bi-2223 \cite{warnes1986critical, mueller2007critical}. However, Warnes~\textit{et al.} also reported an enhanced $\sigma/\mu$ $\sim$ 0.14 for a severely sausaged $\mathrm{Nb-Ti}$ wire at 4 T (reduced field, $H/H_{\text{irr}}$ $\sim$ 0.36) which increased to $\sim$ 0.33 at 10 T \cite{warnes1986critical}, thus correlates a high $\sigma/\mu$ value with strong filament sausaging. Here we measured the $I_{\text{c}}$ distributions of Bi-2212 wires up to 15 T but since this is only 10 – 15\% of $H_{\text{irr}}$, the complete $H/H_{\text{irr}}$ dependence of $\sigma/\mu$ in our wires is not known. Hence, we believe that it is more appropriate to compare the low field $\sigma/\mu$ values of $\mathrm{Nb-Ti}$ to our Bi-2212 wires. The relative broadening width near the irreversibility field seen in the $\mathrm{Nb-Ti}$ probably results from large fluctuations in the effective vortex pinning near $H_{\text{irr}}$ as the elementary pinning forces generated by the dense $\alpha$-Ti vortex-pins become proximity coupled to the superconducting $\beta$-phase matrix\cite{meingast1989quantitative}. 

In line with our original conjecture, we did find a marked difference in $\sigma/\mu$ between the Engi-Mat (pmm180410) and Nexans (pmm100913) wires.  It is evident (see Fig. 2(b)) that the older and lower-$J_{\text{c}}$ Nexans wire has about 40\% higher $\sigma/\mu$ ($\sim$ 0.1 vs. 0.14). As noted earlier by Brown \textit{et al.} \cite{brown2019prediction}, many wires made with Nexans, Engi-Mat and other powder sets had the same normalized $J_{\text{c}}(H)$ characteristics, even though the magnitude of $J_{\text{c}}$ varied by almost 6. We conclude that the effective percolative connectivity of wires is what is really controlling the $J_{\text{c}}$ magnitude. The modern fine particle Engi-Mat powder has more uniform characteristics than the earlier generation Nexans powder as demonstrated experimentally by Jiang \textit{et al.} \cite{jiang2019high}. We conclude that much of the $\sim$ 60\% $J_{\text{c}}$ improvement is due to improved filament connectivity of the Engi-Mat powder \cite{jiang2019high}. Thus, we conclude that the 30\% lower $\sigma/\mu$ in our Engi-Mat sample compared to the lower-$J_{\text{c}}$ Nexans sample is due to improved connectivity associated with the much finer and more uniform Engi-Mat powder. 

As noted above, the Nexans $\sigma/\mu$ values are very similar to that of a severely sausaged $\mathrm{Nb-Ti}$ wire \cite{warnes1986critical}. Knowing that Nexans powder has many hard particles of varying size which distort the filament structure, it is plausible to compare the degraded connectivity of Bi-2212 wire made with Nexans powder to sausaged $\mathrm{Nb-Ti}$ wire \cite{warnes1986critical}. However, we do not conclude that filament sausaging is the sole reason for the degraded $\sigma/\mu$ of our Nexans sample. It is interesting that a Bi-2223 wire ($\mu \sim$ 143.6 A, $\sigma \sim$ 16.8 A, $\sigma/\mu$ $\sim$ 0.12 at 4.2 K, 0.5 T) also had a large $\sigma/\mu$\cite{hornung2010current}. Many other factors are also still in play, one being degraded grain-to-grain connectivity that is certainly present in both Bi-2212 and Bi-2223.

Although better connected than the Nexans powder wire, the higher-$J_{\text{c}}$ Engi-Mat powder wire has a $\sim$ 3 fold higher $\sigma/\mu$, and an almost 50\% smaller $n$ value compared to our high-$J_{\text{c}}$ $\mathrm{Nb-Ti}$ wire. Based on the experimental evidence it can be expected that $I_{\text{c}}$ of Bi-2212 RWs could be increased further if the $\sigma/\mu$ can be minimized by identifying and addressing connectivity limitations. To this end, we believe that the $\sigma/\mu$ value can be used as a quantitative parameter for conductor quality control.

\section{Conclusion}
In fact, the $I_{\text{c}}$ distribution, though wider for Bi-2212 than for optimized Nb--47 wt.\% Ti, is not a great deal larger, in spite of Bi-2212 being strongly anisotropic along each filament, possessing current-blocking regions, and having highly non-uniform filament shapes. The much lower-$J_{\text{c}}$ Nexans wire has the same normalized vortex pinning (same normalized $J_{\text{c}}$($H$) curve) and a broader $\sigma/\mu$, suggestive of poorer connectivity. The relative standard deviation of Engi-Mat wire is $\sim$ 30\% lower than Nexans wire. The implication is that effective filament connectivity can be quantified from the $d^{2}V/dI^2$ measurement. We believe that measurements like these can be a useful tool in further understanding of how best to raise $J_{\text{c}}$ in Bi-2212 conductors.

\section*{Acknowledgment}

We acknowledge the help of G. Bradford, V. S. Griffin, and E. Miller of the National High Magnetic Field Laboratory.

\bibliographystyle{IEEEtran}

\bibliography{References}

\begin{thebibliography}{10}
\providecommand{\url}[1]{#1}
\csname url@samestyle\endcsname
\providecommand{\newblock}{\relax}
\providecommand{\bibinfo}[2]{#2}
\providecommand{\BIBentrySTDinterwordspacing}{\spaceskip=0pt\relax}
\providecommand{\BIBentryALTinterwordstretchfactor}{4}
\providecommand{\BIBentryALTinterwordspacing}{\spaceskip=\fontdimen2\font plus
\BIBentryALTinterwordstretchfactor\fontdimen3\font minus
  \fontdimen4\font\relax}
\providecommand{\BIBforeignlanguage}[2]{{%
\expandafter\ifx\csname l@#1\endcsname\relax
\typeout{** WARNING: IEEEtran.bst: No hyphenation pattern has been}%
\typeout{** loaded for the language `#1'. Using the pattern for}%
\typeout{** the default language instead.}%
\else
\language=\csname l@#1\endcsname
\fi
#2}}
\providecommand{\BIBdecl}{\relax}
\BIBdecl

\bibitem{larbalestier2014isotropic}
D.~C. Larbalestier, J.~Jiang, U.~P. Trociewitz, F.~Kametani, C.~Scheuerlein,
  M.~Dalban-Canassy, M.~Matras, P.~Chen, N.~C. Craig, P.~J. Lee, and E.~E.
  Hellstrom, ``Isotropic round-wire multifilament cuprate superconductor for
  generation of magnetic fields above 30 \uppercase{T},'' \emph{Nature
  Materials}, vol.~13, no.~4, pp. 375--381, 2014.

\bibitem{miao2012recent}
H.~Miao, Y.~Huang, S.~Hong, and J.~A. Parrell, ``Recent advances in
  \uppercase{B}i-2212 round wire performance for high field applications,''
  \emph{IEEE Transactions on Applied Superconductivity}, vol.~23, no.~3, p.
  6400104, 2012.

\bibitem{parrell2004nb3sn}
J.~Parrell, M.~Field, Y.~Zhang, and S.~Hong, ``\uppercase{N}b$_3$\uppercase{S}n
  conductor development for fusion and particle accelerator applications,''
  \emph{AIP Conference Proceedings}, vol. 711, no.~1, pp. 369--375, 2004.

\bibitem{jiang2019high}
J.~Jiang, G.~Bradford, S.~I. Hossain, M.~D. Brown, J.~Cooper, E.~Miller,
  Y.~Huang, H.~Miao, J.~A. Parrell, M.~White \emph{et~al.}, ``High-performance
  \uppercase{B}i-2212 round wires made with recent powders,'' \emph{IEEE
  Transactions on Applied Superconductivity}, vol.~29, no.~5, pp. 1--5, 2019.

\bibitem{kopylov1990role}
V.~Kopylov, A.~Koshelev, I.~Schegolev, and T.~Togonidze, ``The role of surface
  effects in magnetization of high-$\uppercase{T}_{\text{c}}$
  superconductors,'' \emph{Physica C: Superconductivity}, vol. 170, no. 3-4,
  pp. 291--297, 1990.

\bibitem{brown2019prediction}
M.~D. Brown, J.~Jiang, C.~Tarantini, D.~Abraimov, G.~Bradford, J.~Jaroszynski,
  E.~E. Hellstrom, and D.~C. Larbalestier, ``Prediction of the
  $\uppercase{J}_{\text{c}}(\uppercase{B})$ behavior of \uppercase{B}i-2212
  wires at high field,'' \emph{IEEE Transactions on Applied Superconductivity},
  vol.~29, no.~5, pp. 1--4, 2019.

\bibitem{kametani2015comparison}
F.~Kametani, J.~Jiang, M.~Matras, D.~Abraimov, E.~Hellstrom, and
  D.~Larbalestier, ``Comparison of growth texture in round \uppercase{B}i2212
  and flat \uppercase{B}i2223 wires and its relation to high critical current
  density development,'' \emph{Scientific Reports}, vol.~5, p. 8285, 2015.

\bibitem{shen2011heat}
T.~Shen, J.~Jiang, F.~Kametani, U.~P. Trociewitz, D.~C. Larbalestier, and E.~E.
  Hellstrom, ``Heat treatment control of
  \uppercase{A}g--\uppercase{B}i$_2$\uppercase{S}r$_2$\uppercase{C}a\uppercase{C}u$_2$\uppercase{O}$_\text{x}$
  multifilamentary round wire: investigation of time in the melt,''
  \emph{Superconductor Science and Technology}, vol.~24, no.~11, p. 115009,
  2011.

\bibitem{jiang2016effects}
J.~Jiang, A.~Francis, R.~Alicea, M.~Matras, F.~Kametani, U.~P. Trociewitz,
  E.~E. Hellstrom, and D.~C. Larbalestier, ``Effects of filament size on
  critical current density in overpressure processed \uppercase{B}i-2212 round
  wire,'' \emph{IEEE Transactions on Applied Superconductivity}, vol.~27,
  no.~4, pp. 1--4, 2016.

\bibitem{BAIXERAS19671541}
J.~Baixeras and G.~Fournet, ``Vortex displacement losses in a non-ideal type
  \uppercase{II} superconductor,'' \emph{Journal of Physics and Chemistry of
  Solids}, vol.~28, no.~8, pp. 1541--1547, 1967.

\bibitem{warnes1986critical}
W.~Warnes and D.~Larbalestier, ``Critical current distributions in
  superconducting composites,'' \emph{Cryogenics}, vol.~26, no.~12, pp.
  643--653, 1986.

\bibitem{warnes1988model}
W.~H. Warnes, ``A model for the resistive critical current transition in
  composite superconductors,'' \emph{Journal of Applied Physics}, vol.~63,
  no.~5, pp. 1651--1662, 1988.

\bibitem{mueller2007critical}
H.~Mueller, F.~Hornung, A.~Rimikis, and T.~Schneider, ``Critical current
  distribution in composite superconductors,'' \emph{IEEE Transactions on
  Applied Superconductivity}, vol.~17, no.~2, pp. 3757--3760, 2007.

\bibitem{li2015rrr}
P.~Li, L.~Ye, J.~Jiang, and T.~Shen, ``\uppercase{RRR} and thermal conductivity
  of \uppercase{A}g and \uppercase{A}g-0.2 wt.\% \uppercase{M}g alloy in
  \uppercase{A}g/\uppercase{B}i-2212 wires,'' in \emph{IOP Conference Series:
  Materials Science and Engineering}, vol. 102, no.~1.\hskip 1em plus 0.5em
  minus 0.4em\relax IOP Publishing, 2015, p. 012027.

\bibitem{bonura2018very}
M.~Bonura, F.~Avitabile, C.~Barth, J.~Jiang, D.~Larbalestier, A.~F{\^e}te,
  A.~Leo, L.~Bottura, and C.~Senatore, ``Very-high thermal and electrical
  conductivity in overpressure-processed
  \uppercase{B}i$_2$\uppercase{S}r$_2$\uppercase{C}a\uppercase{C}u$_2$\uppercase{O}$_\text{8+x}$
  wires,'' \emph{Materials Research Express}, vol.~5, no.~5, p. 056001, 2018.

\bibitem{bruzzone1996bench}
P.~Bruzzone, H.~H. ten Kate, M.~Nishi, A.~Shikov, J.~Minervini, and
  M.~Takayasu, ``Bench mark testing of \uppercase{N}b$_3$\uppercase{S}n strands
  for the \uppercase{ITER} model coil,'' in \emph{Advances in Cryogenic
  Engineering Materials}, L.~T. Summers, Ed.\hskip 1em plus 0.5em minus
  0.4em\relax Boston, MA: Springer, US, 1996, pp. 1351--1358.

\bibitem{shen2019stable}
T.~Shen, E.~Bosque, D.~Davis, J.~Jiang, M.~White, K.~Zhang, H.~Higley,
  M.~Turqueti, Y.~Huang, H.~Miao \emph{et~al.}, ``Stable, predictable and
  training-free operation of superconducting \uppercase{B}i-2212 rutherford
  cable racetrack coils at the wire current density of 1000
  \uppercase{A}/mm$^2$,'' \emph{Scientific Reports}, vol.~9, no.~1, pp. 1--9,
  2019.

\bibitem{plummer1987dependence}
C.~Plummer and J.~Evetts, ``Dependence of the shape of the resistive transition
  on composite inhomogeneity in multifilamentary wires,'' \emph{IEEE
  Transactions on Magnetics}, vol.~23, no.~2, pp. 1179--1182, 1987.

\bibitem{kimmich1999investigation}
R.~Kimmich, A.~Rimikis, and T.~Schneider, ``Investigation of critical current
  distribution in composite superconductors,'' \emph{IEEE Transactions on
  Applied Superconductivity}, vol.~9, no.~2, pp. 1759--1762, 1999.

\bibitem{savitzky1964smoothing}
A.~Savitzky and M.~J. Golay, ``Smoothing and differentiation of data by
  simplified least squares procedures.'' \emph{Analytical Chemistry}, vol.~36,
  no.~8, pp. 1627--1639, 1964.

\bibitem{edelman1993resistive}
H.~S. Edelman and D.~C. Larbalestier, ``Resistive transitions and the origin of
  the $n$ value in superconductors with a gaussian critical-current
  distribution,'' \emph{Journal of Applied Physics}, vol.~74, no.~5, pp.
  3312--3315, 1993.

\bibitem{hampshire1987detailed}
D.~Hampshire and H.~Jones, ``A detailed investigation of the $\uppercase{E} -
  \uppercase{J}$ characteristic and the role of defect motion within the
  flux-line lattice for high-current-density, high-field superconducting
  compounds with particular reference to data on
  \uppercase{N}b$_3$\uppercase{S}n throughout its entire field-temperature
  phase space,'' \emph{Journal of Physics C: Solid State Physics}, vol.~20,
  no.~23, p. 3533, 1987.

\bibitem{cai1998current}
X.~Cai, A.~Polyanskii, Q.~Li, G.~Riley, and D.~Larbalestier, ``Current-limiting
  mechanisms in individual filaments extracted from superconducting tapes,''
  \emph{Nature}, vol. 392, no. 6679, pp. 906--909, 1998.

\bibitem{jones1967non}
R.~Jones, E.~Rhoderick, and A.~Rose-Innes, ``Non-linearity in the
  voltage-current characteristic of a type-2 superconductor,'' \emph{Physics
  Letters A}, vol.~24, no.~6, pp. 318--319, 1967.

\bibitem{suenaga1991irreversibility}
M.~Suenaga, A.~Ghosh, Y.~Xu, and D.~Welch, ``Irreversibility temperatures of
  \uppercase{N}b$_3$\uppercase{S}n and \uppercase{N}b-\uppercase{T}i,''
  \emph{Physical Review Letters}, vol.~66, no.~13, p. 1777, 1991.

\bibitem{brandt1995flux}
E.~H. Brandt, ``The flux-line lattice in superconductors,'' \emph{Reports on
  Progress in Physics}, vol.~58, no.~11, p. 1465, 1995.

\bibitem{hornung2010current}
F.~Hornung, A.~Rimikis, and T.~Schneider, ``Current sharing and critical
  current distribution in \uppercase{B}i-2223 tapes,'' \emph{IEEE Transactions
  on Applied Superconductivity}, vol.~20, no.~3, pp. 1589--1592, 2010.

\bibitem{meingast1989quantitative}
C.~Meingast, P.~Lee, and D.~Larbalestier, ``Quantitative description of a high
  $\uppercase{J}_{\text{c}}$ \uppercase{N}b-\uppercase{T}i superconductor
  during its final optimization strain. \uppercase{I}. microstructure,
  $\uppercase{T}_{\text{c}}$, $\uppercase{H}_{\text{c2}}$, and resistivity,''
  \emph{Journal of Applied Physics}, vol.~66, no.~12, pp. 5962--5970, 1989.

\end{thebibliography}
\end{document}